\def\keV{{\rm keV}}

\documentstyle[11pt,aaspp4,epsf]{article}
\begin{document}
\slugcomment{Published in {\em Nature} 22 November 2001}

\begin{center}
{\bf Regulation of the X-ray luminosity of clusters of 
     galaxies by cooling \\ and supernova feedback }
\end{center}
\author{G. Mark Voit$^*$ \& Greg L. Bryan$^{\dag \ddag}$} 
\vspace*{1em}
\affil{$^*$STScI, 3700 San Martin Drive, Baltimore, MD 21218, USA}
\affil{$^\dag$Department of Physics, MIT, Cambridge, MA 02139, USA}
\affil{$^\ddag$Oxford University, Astrophysics, Keble Road, Oxford, OX1 3RH, UK}


{\bf

Clusters of galaxies are thought to contain about ten times as much
dark matter as baryonic matter$^1$.  The dark component therefore
dominates the gravitational potential of a cluster, and the baryons
confined by this potential radiate X-rays with a luminosity that 
depends mainly on the gas density in the cluster's core.$^2$ 
Predictions of the X-rays' properties based on models of cluster 
formation do not, however, agree with the observations.  If the
models ignore the condensation of cooling gas into stars and 
feedback from the associated supernovae, they overestimate the 
X-ray luminosity because the core gas is too high.  An early 
episode of uniformly distributed supernova feedback could rectify 
this by heating the uncondensed gas and therefore making it harder 
to compress into the core$^{3-11}$, but such a process seems to
require an implausibly large number of supernovae$^{6,8,12-14}$.  
Here we show how radiative cooling of intergalactic gas and subsequent 
supernova heating conspire to eliminate highly compressible 
low-entropy gas from the intracluster medium.  This brings 
the core entropy and X-ray luminosities of galaxy clusters into 
agreement with the observations, in a way that depends little on
the efficiency of supernova heating in the early Universe. 

}

\vspace*{1em}

Numerical simulations of cosmological structure formation show that
the dark-matter potential wells of clusters are quite similar in
shape across a wide mass range$^{15}$.  A cluster's mean mass density
within the virial radius $r_v$ is roughly 200 times $\rho_{cr}$, 
the critical density for closing the universe, with a dark-matter 
density distribution $\rho_{DM} \propto [r(1+cr)^2]^{-1}$, where $r$ 
is the radius in units of $r_v$ and $c \sim 4-10$ is a parameter 
governing the concentration of dark matter toward the centre of 
the cluster.  In the absence of radiative cooling and supernova 
heating, the gas density in such simulated clusters follows nearly 
the same profile$^{16}$, with some additional flattening 
for $r \lesssim 1/c$ .

If all clusters had similar gas density distributions, then cluster 
temperature would scale as $T \propto \rho_{cr} r_v^2$ and 
X-ray luminosity would scale 
as $L \propto \Lambda \rho_{cr}^2 r_v^3$, where $\Lambda$ 
is the cooling rate per unit density of the hot gas. At typical cluster 
temperatures ($T > 2 \, \keV$), free-free emission dominates the cooling, 
so $\Lambda \propto T^{1/2}$ and we would expect $L \propto T^2$.  
Instead, the observed power-law index$^{17-19}$ in the relation 
$L \propto T^b$ is $b \approx 2.6-2.9$, meaning that low-temperature 
clusters and groups of galaxies are far less luminous 
than expected.  This luminosity deficit is also evident in the low level 
of the unresolved $\sim 1 \, \keV$ X-ray background, which is an order 
of magnitude smaller than predicted by cosmological simulations without 
radiative cooling or supernova heating$^{20-23}$.  

The X-ray luminosities of low-temperature clusters are unexpectedly small
because their gas is less centrally concentrated than in hotter clusters, 
an effect that has been attributed to a universal minimum entropy level 
in intracluster gas resulting either from supernova heating$^{5,9-11}$,
from heating by active galactic nuclei$^8$, or from radiative 
cooling$^{8,12,24,25}$.  A lower limit to intracluster entropy steepens 
the relationship between luminosity and temperature because the shallower 
potential wells of low-temperature clusters are less able to overcome 
the resistance to compression owing to that minimum entropy. 
Both observations$^5$ and theoretical models$^{6,10,11,20,21}$ have 
established that a core entropy level corresponding to $S \sim 100-200 
\, \keV \, {\rm cm}^2$ can account for the observed slope of the 
$L$:$T$ relation and the low level of the $\sim 1 \, \keV$ background.

This entropy scale emerges most naturally from considerations involving 
radiative cooling.  For any cluster of temperature $T$, one can compute
the specific entropy level at which a gas parcel would radiate away its 
thermal energy in a time equivalent to the age of the Universe.  
Figure~1 shows that locus in the entropy-temperature plane for a 
typical cluster heavy-element abundance of 30\% of the solar value.  
The locus for 
this typical abundance lies in the $100-150 \, {\rm keV \, cm^2}$ range for 
$T < 2 \, \keV$ and rises in proportion to $T^{2/3}$ at higher temperatures.  
Note that the measured core entropies of clusters and groups$^5$ 
closely track the cooling locus.

The remarkable correspondence between the threshold 
entropy for cooling ($S_c$) and observations of core entropy 
suggests the following picture for the evolution of the intracluster
medium.  We can consider the medium to consist of
independent gas parcels, each with its own entropy history.
The luminosity of a sufficiently relaxed cluster, in which 
entropy increases monotonically with radius, is then completely
determined by the shape of its potential well $\rho_{DM}(r)$ and
the distribution of specific entropy among its gas parcels.  
In the context of hierarchical merging, a parcel's entropy history
proceeds like this: each merger event affecting the parcel raises
its entropy to some new value $S_i$, and as the gas relaxes 
after the merger, the parcel's temperature will approach $T_i$, 
the characteristic temperature of its new dark-matter halo.  
A parcel's trajectory in the $S-T$ plane is thus defined by a 
series of points $(S_i,T_i)$ that proceed upward and to the 
right in a diagram like Figure 1.

Now consider the effect of the cooling locus, which creeps vertically
up the diagram as time progresses.  If a parcel's entropy trajectory
always remains above this threshold, it will never cool.  However, parcels
of gas that find themselves below the cooling threshold as structure develops 
begin to condense and form stars.  Shortly thereafter, supernovae 
heat the neighbouring gas with an efficiency that remains unknown,
adding entropy that tends to suppress further cooling and condensation.
Stars will continue to form and supernovae will continue to explode
until there is no more gas below the cooling threshold; then both processes
must cease.  Thus cooling and whatever feedback accompanies it 
act in tandem to eliminate gas parcels with $S < S_c$, inevitably 
creating an entropy floor at the cooling threshold.

This model leads to an $L$:$T$ relation that closely matches 
cluster observations.  Suppose that both the gas and dark matter
density are proportional to $[r (1+cr)^2]^{-1}$ in the absence of cooling.
Then we can compute the unmodified entropy distribution $M_g(S)$, 
the mass of gas with specific entropy less than $S$, and the inverse relation 
$S(M_g)$.  When cooling is allowed to operate, gas parcels with
$S < S_c$ are subject to condensation, and any subsequent feedback 
is targeted directly at those low-entropy gas parcels.  If feedback
is inefficient, condensation will remove this gas from the intracluster
medium.  If feedback is highly efficient, supernova heating will
raise the entropy of this gas to $S \gg S_c$, and it will convect to
the outer regions of the cluster.  In either limit, the modified 
entropy distribution approaches $S^*(M_g) = S(M_g + M_c)$, where
$M_c \equiv M_g(S_c)$.  Figure~2 illustrates the $L$:$T$ relation 
that we compute when we allow gas with entropy distribution $S^*(M_g)$ 
to be in hydrostatic equilibrium with the dark matter potential, 
assuming a Hubble constant of $100\, h \, {\rm km \, s^{-1} \, 
Mpc^{-1}}$ with $h = 0.65$ and a baryon fraction $0.02 \, h^{-2} 
\Omega_M^{-1}$ with a total matter density $\Omega_M = 0.33$.  
(Using an alternative modified 
entropy distribution, $S^*(M_g) = S(M_g) + S_c$, leads to similar 
results.)

The $L$:$T$ relation we derive from the cooling threshold agrees 
not only with observations but also with cluster simulations$^{26}$ 
that include radiative cooling but no supernova heating.  
That agreement underlines an important feature of this model: 
the $L$:$T$ relation it predicts is insensitive to the
efficiency of supernova heating.  Any combination of cooling
and feedback that truncates the unmodifed entropy distribution
at the entropy scale $S_c$ will lead to a similar $L$:$T$
relation.  In that sense, the relation determined by
the cooling threshold represents the upper envelope to
all $L$:$T$ relations consistent with the age of the universe.
Extreme amounts of feedback could further lower the luminosity,
but apart from relatively small amounts of low-entropy gas
that have not yet completely cooled$^{27}$, the luminosity
of a cluster cannot substantially exceed this relation.

Despite the spherical symmetry of the end state, the cooling,
condensation, and feedback processes that regulate the $L$:$T$
relation are not restricted to the cluster's core.  In hierarchical
models of structure formation, the spatial distribution of specific
entropy is initially quite complex.  If cooling is not allowed to 
operate, the highest-density, lowest-entropy gas parcels will eventually
find their way into the cluster core.  However, in the real Universe,
this low-entropy gas condenses and forms galaxies long before it 
reaches the centre of the cluster.  Much of the condensation and 
feedback that ultimately determines a cluster's luminosity therefore 
pre-dates the epoch of cluster formation.  Only modest quantities of gas
remain below the cooling threshold in most present-day clusters, 
where they collect into centrally-focused cooling flows. 

If this interpretation of the $L$:$T$ relation is correct 
then the level of the entropy floor in clusters ($\sim 100-200 
\, \keV \, {\rm cm}^2$) conveys little information about
the efficiency of supernova heating in the early Universe. 
That information is more directly reflected in the fraction
of baryons that have formed stars.  A large proportion of
the baryons that are now in clusters were associated with
objects of less than $10^{12}$ solar masses at $z \approx 2$.
At that time, the cooling threshold in a dark-matter halo of
$10^{12}$ solar masses would have been $\sim 10 \, \keV \, {\rm cm}^2$, 
and virtually all the baryons in those haloes were below that
threshold$^{28}$.  Some form of feedback, probably supernovae, 
prevented most of these baryons from condensing.  Once 
feedback drives uncondensed gas to high entropy and 
low density, merger shocks can more easily maintain it above 
the cooling threshold.  Detailed computations tracing
the entropy history of each gas parcel through the series of 
cooling, feedback, and merger events leading to the final
cluster will therefore be needed to relate the supernova
heating efficiency to the fraction of baryons now in stars.


\vspace*{2em}

Received 2 July; accepted 26 September 2001.

\vspace*{1em}


1.  Evrard, A. E.  
    The intracluster gas fraction in X-ray clusters: constraints
    on the clustered mass density. 
    {\em Mont. Not. R. Astron. Soc.}, {\bf 292}, 289-297 (1997).

2.  Sarazin, C. L.
    {\em X-ray emissions from clusters of galaxies}
    (Cambridge University Press, Cambridge, 1988)
    
3.  Evrard, A. E. \& Henry, J. P. 
    Expectations for X-ray cluster observations by the ROSAT satellite.  
    {\em Astrophys. J.}, {\bf 383}, 95-103 (1991). 

4.  Kaiser, N. 
    Evolution of clusters of galaxies. 
    {\em Astrophys. J.}, {\bf 383}, 104-111 (1991).

5.  Ponman, T. J., Cannon, D. B., \& Navarro, J. F. 
    The thermal imprint of galaxy formation on X-ray clusters. 
    {\em Nature}, {\bf 397}, 135-137 (1999).

6.  Balogh, M. L., Babul, A., \& Patton, D. R. 
    Pre-heated isentropic gas in groups of galaxies. 
    {\em Mont. Not. R. Astron. Soc.}, {\bf 307}, 463-479 (1999).

7.  Wu, K. K. S., Fabian, A. C., \& Nulsen, P. E. J. 
    The effect of supernova heating on cluster properties and constraints
    on galaxy formation models. 
    {\em Mont. Not. R. Astron. Soc.}, {\bf 301}, L20-L24 (1998).

8.  Wu, K. K. S., Fabian, A. C., \& Nulsen, P. E. J. 
    Non-gravitational heating in the hierarchical formation of X-ray clusters. 
    {\em Mont. Not. R. Astron. Soc.}, {\bf 318}, 889-912 (2000).

9.  Cavaliere, A., Menci, N., \& Tozzi, P.
    Hot gas in clusters of galaxies: the punctuated equilibria model.
    {\em Mont. Not. R. Astron. Soc.}, {\bf 308}, 599-608 (1999).

10. Tozzi, P. \& Norman, C.  
    The evolution of X-ray clusters and the entropy of the intracluster
    medium.
    {\em Astrophys. J.}, {\bf 546}, 63-84 (2001).

11. Bialek, J. J., Evrard, A. E., \& Mohr, J. J.
    Effects of preheating on X-ray scaling relations in galaxy clusters.
    {\em Astrophys. J.}, {\bf 555}, 597-612 (2001).

12. Bryan, G. L. 
    Explaining the entropy excess in clusters and groups of galaxies
    without additional heating.
    {\em Astrophys. J.}, {\bf 544}, L1-L5 (2001).

13. Kravtsov, A. V. \& Yepes, G. 
    On the supernova heating of the intergalactic medium. 
    {\em Mont. Not. R. Astron. Soc.}, {\bf 318}, 227-238 (2000).

14. Bower, R. et al.
    The impact of galaxy formation on the X-ray evolution of clusters.
    {\em Mont. Not. R. Astron. Soc.},  {\bf 325}, 497-508 (2001).    

15. Navarro, J. F., Frenk, C. S., \& White, S. D. M. 
    A universal density profile from hierarchical clustering. 
    {\em Astrophys. J.}, {\bf 490}, 493-508 (1997).

16. Frenk, C. S. et al.
    The Santa Barbara cluster comparison project: a comparison of
    cosmological hydrodynamics simulations.
    {\em Astrophys. J.}, {\bf 525}, 554-582 (1999).

17. Edge, A. C. \& Stewart, G. C.
    EXOSAT observations of clusters of galaxies. I - The X-ray data.
    {\em Mont. Not. R. Astron. Soc.}, {\bf 252}, 414-427 (1991).

18. Markevitch, M.
    The $L_X$-$T$ relation and temperature function for nearby clusters
    revisited.
    {\em Astrophys. J.}, {\bf 504}, 27-34 (1998).

19. Arnaud, M. \& Evrard, A. E. 
    The $L_X$-$T$ relation and intracluster gas fractions of X-ray clusters. 
    {\em Mont. Not. R. Astron. Soc.}, {\bf 305}, 631-640 (1999).

20. Pen, U.
    Heating of the intergalactic medium.
    {\em Astrophys. J.}, {\bf 510}, L1-L5 (1999).

21. Voit, G. M. \& Bryan, G. L.  
    On the distribution of X-ray surface brightness from diffuse gas.
    {\em Astrophys. J.}, {\bf 551}, L139-L142 (2001).

22. Bryan, G. L. \& Voit, G. M. 
    The X-ray surface brightness distribution from diffuse gas.
    {\em Astrophys. J.}, {\bf 556}, 590-600 (2001).

23. Wu, K. K. S., Fabian, A. C., \& Nulsen, P. E. J. 
    The soft X-ray background: evidence for widespread disruption
    of the gas halos of galaxy groups. 
    {\em Mont. Not. R. Astron. Soc.}, {\bf 324}, 95-107 (2001).

24. Knight, P. A. \& Ponman, T. J.
    The properties of the hot gas in galaxy groups and clusters from
    1D hydrodynamical simulations - I. Cosmological infall models.
    {\em Mont. Not. R. Astron. Soc.}, {\bf 289}, 955-972 (1997).

25. Pearce, F. R., Thomas, P. A., Couchman, H. M. P., \& Edge, A. C.
    The effect of radiative cooling on the X-ray properties of
    galaxy clusters. 
    {\em Mont. Not. R. Astron. Soc.}, {\bf 317}, 1029-1049 (2000).

26. Muanwong, O., Thomas, P. A., Kay, S. T., 
    Pearce, F. R., \& Couchman, H. M. P. 
    The effect of radiative cooling on scaling laws of clusters and groups. 
    {\em Astrophys. J.}, {\bf 552}, L27-L30 (2001).

27. Fabian, A. C., Crawford, C. S., Edge, A. C., Mushotsky, R. F.
    Cooling flows and the X-ray luminosity-temperature relation for clusters.
    {\em Mont. Not. R. Astron. Soc.}, {\bf 267}, 779-784 (1994).

28. Balogh, M., Pearce, F. R., Bower, R. G., \& Kay, S. T.
    Revisiting the cosmic cooling crisis.
    {\em Mont. Not. R. Astron. Soc.} {\bf 326}, 1228-1234  (2001)    

29. Eke, V.,Navarro, J. F., \& Steinmetz, M.
    The power spectrum dependence of dark matter halo concentrations.
    {\em Astrophys. J.}, {\bf 554}, 114-125 (2001).

30. Helsdon, S. F. \& Ponman, T. J.
    The intragroup medium in loose groups of galaxies.
    {\em Mont. Not. R. Astron. Soc.}, {\bf 315}, 356-370 (2000).

{\bf Acknowledgements.} 
We acknowledge R. Bower and M. Balogh for discussions.  
G.M.V. receives partial support from NASA, and G.L.B. 
was supported by NASA through a Hubble Fellowship.

\newpage

\begin{figure}
\plotone{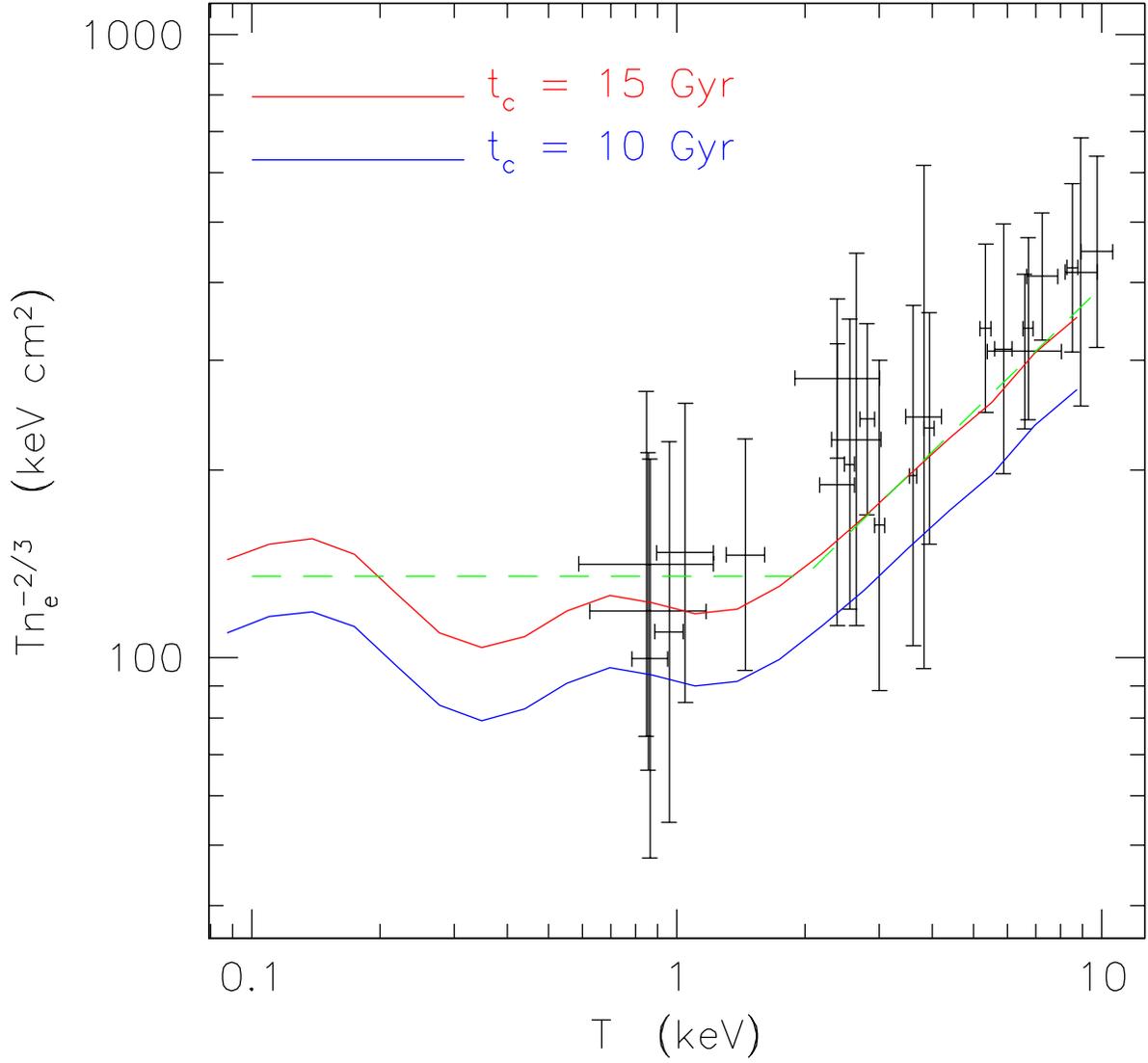}
\figcaption[centro_pon.ps]{
Threshold entropy for cooling within
the age of the universe.  The solid lines show the specific
entropy $S = T n_e^{-2/3}$, where $n_e$ is the electron density,
at which gas of temperature $T$ will cool within 15 Gyr (red) 
and 10 Gyr (blue) for elemental abundances of 30\% solar.  
(The usual thermodynamic entropy for an ideal monatomic gas 
is proportional to $\ln S$.) Note that the core entropy levels
in clusters measured at the core radius $0.1 \, r_v$ (ref. 5)
track the cooling threshold for typical 
cluster heavy-element abundances of 30\% solar.  Within that
core radius many clusters contain gas whose entropy lies
below the cooling threshold, but the luminosity of that relatively
small amount of low-entropy gas usually contributes only 
modestly to the overall luminosity of the cluster.$^{18}$
The dashed green line indicates the entropy threshold $S_c$ 
used to determine the $L$:$T$ relation in Figure~2. 
}
{\label{centro_pon}}
\end{figure}

\begin{figure}
\plotone{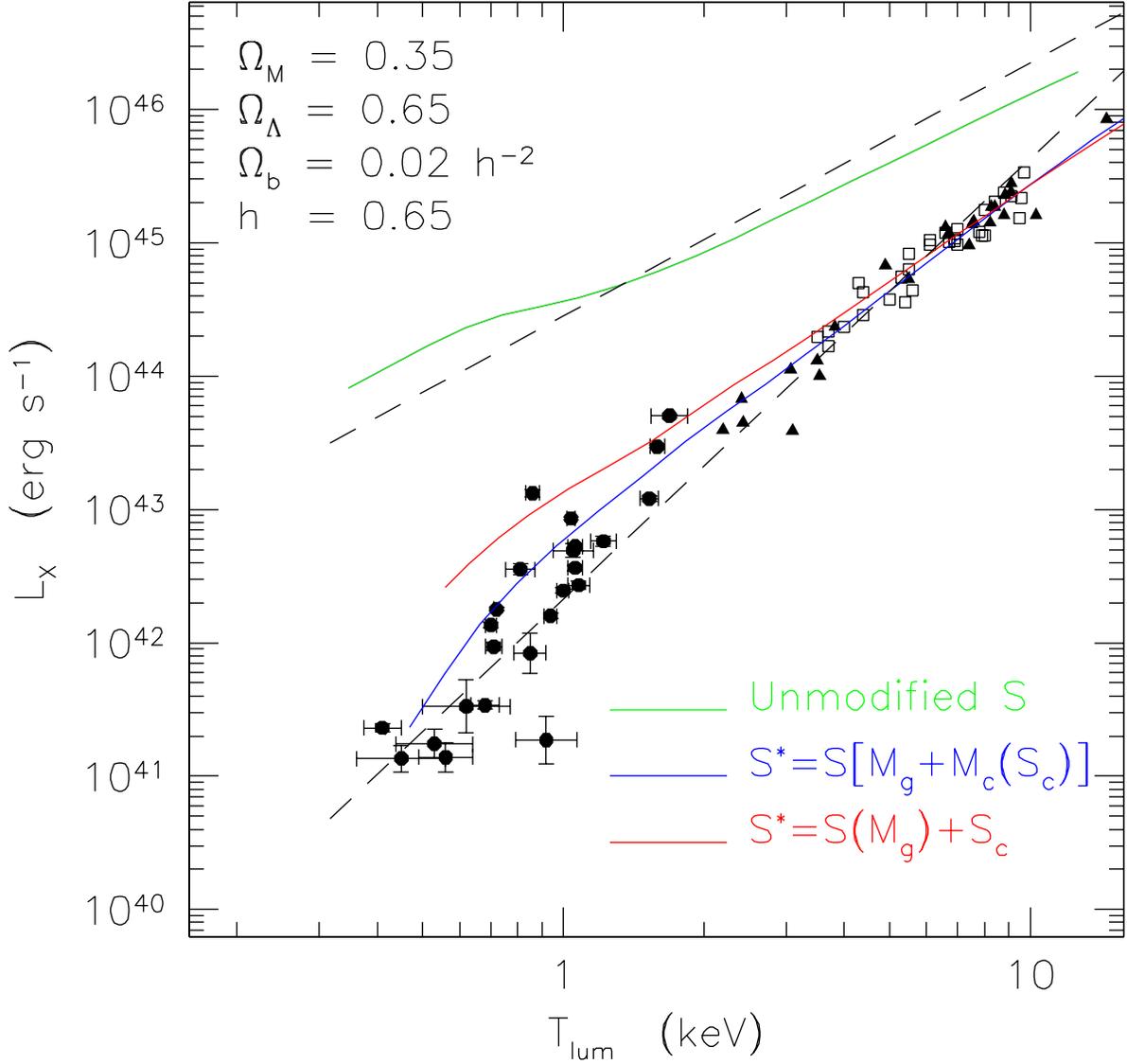}
\figcaption[ltplot.ps]{
Relation between bolometric X-ray luminosity $L_{X}$
and luminosity-weighted temperature $T_{lum}$ in clusters and groups
of galaxies.  Green line, the $L$:$T$ relation from the
unmodified entropy distribution $S(M_g)$ for concentration factors
corresponding to model $S_{1.2}$ from ref. 29.  
Blue line, the $L$:$T$ relation derived from our 
primary cooling-threshold model, in which $S^* = S[M_g + M_c(S_c)]$ 
(see text for details) for the threshold entropy $S_c$
indicated by the dashed green line in Figure~1.  Red line,
the $L$:$T$ relation for an alternative modified-entropy model
in which $S^* = S(M_g) + S_c$.  Dashed lines, fits 
to numerical simulations$^{25}$ of clusters with (lower line) 
and without (upper line) radiative cooling.  Triangles,  
measurements of $L_{X}$ and $T$ from a compilation of cluster data$^{19}$;
open squares, additional cluster data$^{18}$; 
filled circles, group data$^{30}$.
All data and models are normalized to a Hubble constant of $h=0.65$.
The cooling-threshold model naturally accounts for the $L$:$T$ relation
in a standard $\Lambda$-dominated cosmology (total matter density, 
$\Omega_M = 0.35$; dark-energy density, $\Omega_\Lambda = 0.65$; 
baryon density, $\Omega_b = 0.02 h^{-2}$) without any adjustable 
parameters.
}
{\label{ltplot}}
\end{figure}




\end{document}